# On the invariant formulation of fluid mechanics

S.Piekarski
IPPT PAN

Abstract

It can be observed that the differential operators of fluid mechanics can be defined in terms of the complete derivative on the finite – dimensional affine space. It follows from the fact that all norms on the finite – dimensional vector space are equivalent and from the definition of the complete derivative on the normed affine spaces (see: L.Schwartz, Analyse Mathematique, Hermann, 1967).
In particular, it is shown that the "substantial derivative" of the standard formulation is a directional derivative along the "non – relativistic four – velocity".

Introduction

Some results on the subject of the invariant formulation of continuum mechanics are given in $[1,2]$. As we shall see, the formulation given in the present text is much simpler. For the convenience of the reader, some facts concerning the structure of the Galilean space – time are briefly recalled below (compare $[1,2]$).
By a Galilean space – time we mean an ordered quadruple ($G, T_G, \gamma, \bullet$) where ($G, T_G$) is a four – dimensional real affine space ($G$ is a set of points of affine space and $T_G$ is a corresponding translation space), $\gamma$ is a non – zero form on $T_G$ (that is, $\gamma \in T_G^*$, where $T_G^*$ is a vector space dual to $T_G$), and "$\bullet$" is a scalar product in the vector space $S$, where

$$S = \{ y \in T_G ; <\gamma, y> = 0 \}. \qquad (1)$$

The absolute time instant $\Delta(g, g')$ between the points $g, g'$ of the Galilean space – time $G$ is defined in the terms of the form $\gamma$ :

$$g, g' \to \Delta \left\langle \gamma, \vec{g - g'} \right\rangle, \qquad (2)$$

where the operation of assigning to the pair $g, g'$ of points from $G$ the vector $\vec{g - g'}$ from $T_G$ is well – defined since the Galilean space – time is the four – dimensional affine space and the operation of "subtraction of points" is allowed by the axioms of the affine space. In the rest of this text, we shall omit the arrow over the subtraction of points of the affine space. Hyperplanes in $G$, composed of simultaneous events, are given by the equivalence classes of the relation

$$g, g', g \sim g' \Leftrightarrow \langle \gamma, g - g' \rangle = 0. \tag{3}$$

The equivalence class represented by $g \in G$ shall be denoted $[g]$. The set $[G] = \{[g], g \in G\}$ can be identified with the set of absolute time instants. By definition, $[G]$ is a one – dimensional affine space and the choice of $\gamma$ endows this space with an orientation (the alternative orientation corresponds to the choice $-\gamma$ instead of $\gamma$).

.The vector space $S$, defined by (1), is a subspace of $T_G$; according to the rules of affine geometry, vectors from $S$ can be added to the points of $G$, and one can see that such action is consistent with (3) in the sense that for every $g \in G$ and for every $s \in S$, $g$ and g+s are equivalent.

The important definition is the definition of a "non–relativistic four-velocity"; it can be observed that the set of all admissible four-velocities can be defined in the terms of the chronological form $\gamma$ by the formula

$$W = \{w \in T_G; \langle \gamma, w \rangle = 1\} \tag{4}$$

(see [1], p.772, the differential operators for a finite – dimensional affine space with a chronological form are discussed in [2]). Now we can define the set of world – lines of inertial observers; they are the straight lines described parametrically as

$$R \ni t \to g + tw \subset G, \tag{5}$$

where $g \in G$ and $w \in W$ but the set of straight lines is smaller since the different parameterizations of the type (5) can describe the same inertial observer. Therefore it is convenient to impose the restriction that $g$ in (5) belongs to a fixed space of simultaneous events, which we shall denote $H_0$; then

$$R \ni t \to g_0 + tw \in G, \quad g_0 \in H_0, w \in W, \tag{6}$$

parameterizes the set of inertial observers bijectively. It is commonly known that the inertial observers parameterize the space – time with the "inertial coordinate systems" and the explicit form of such "inertial coordinate system" in the language of affine geometry is

$$R \times S \ni (t, r) \to g_0 + tw + r \in H_t \subset G. \tag{7}$$

By a "world line" of the material point we shall mean a continuous curve in $G$, parameterized bijectively by the absolute time

$$R \ni t \to \xi(t) \in G. \tag{8}$$

In principle, in order to define the notion of a continuous curve one has to define the notion of continuity, but in our case the curve lies in the Galilean space – time, which is a finite – dimensional affine space and therefore carries a canonical metrizable topology (obviously, this topology is consistent with the "manifold topology", existing on the finite – dimensional affine space as a "finite – dimensional manifold").

Our aim now is to differentiate $\xi(t)$ with respect to the absolute time; however, the problem of the choice of the "derivative" is not quite trivial. The standard approach is to use the differential geometry (as described, for example, in Wintgen – Sulanke [3]). In turn, in [2] we used the dual approach of Peradzyński (compare [4,5]). As it has been observed already in the Abstract, the most natural definition of differential operators on the Galilean space – time $G$ is that basing on the complete derivative and the Schwartz's formalism [6]. However, in this text, we shall use the "affine differential quotient", which in well – defined on the finite – dimensional affine spaces as a consequence of the fact that all norms on the finite – dimensional vector space are equivalent. A detail discussion of the relation between these approaches shall be given in a separate paper.

Now, we define the "tangent vector" to the world line, parameterized by the absolute time, simply as

$$\frac{d}{dt}\xi(t) = \lim_{t' \to t} \frac{\xi(t') - \xi(t)}{t' - t}. \tag{9}$$

The "four – dimensional" definition $(8)$ of the world line $\xi(t)$ can be written equivalently in any inertial coordinate systems $(7)$ as

$$R \times S \ni (t, r(t)) \to g_0 + tw + r(t) \in H_t \subset G. \tag{10}$$

In $(10)$, the radius – vector $r(t)$ depends on the choice of the four – velocity of the corresponding inertial observer. In order to make our notation more rigorous, this dependence can be explicitly taken into account and then $(10)$ assumes the form:

$$R \times S \ni (t, r_w(t)) \to g_0 + tw + r_w(t) \in H_t \subset G. \tag{11}$$

After inserting $(11)$ into $(9)$ one arrives at

$$\frac{d}{dt}\xi(t) = \lim_{t' \to t} \frac{[g_0 + t'w + r_w(t')] - [g_0 + tw + r_w(t)]}{t' - t} = w + \lim_{t' \to t} \frac{r_w(t') - r_w(t)}{t' - t} = w + \frac{d}{dt}r_w(t). \tag{12}$$

In the above expression, $\frac{d}{dt}r_w(t)$ is the "standard" velocity of a material point. It is spatial since $r_w(t)$ is spatial.

The four – velocity of a given material point can be represented in the two different inertial coordinate systems, and then

$$\frac{d}{dt}\xi(t) = w + \frac{d}{dt}r_w(t) = w' + \frac{d}{dt}r_{w'}(t). \tag{13}$$

This identity implies

$$\frac{d}{dt}r_w(t) = [w' - w] + \frac{d}{dt}r_{w'}(t). \tag{14}$$

In order to show that $(14)$ is consistent with the standard "non – relativistic" rule of adding velocities" we have to show that the difference of four - velocities is a spatial vector. However, it is a direct consequence of the definition of a four – velocity, since for two different four – velocities $w$ and $w'$, for which, by definition

$$\langle \gamma, w \rangle = 1, \quad \langle \gamma, w' \rangle = 1, \tag{15}$$

subtraction gives

$$\langle \gamma, w \rangle - \langle \gamma, w' \rangle = \langle \gamma, w - w' \rangle = 0. \tag{16}$$

But, on account of the definition $(1)$, $(16)$ means that $w - w'$ belongs to $S$ and therefore is spatial.
For

$$u_w = \frac{d}{dt} r_w(t), \quad u_{w'} = \frac{d}{dt} r_{w'}(t)$$

the identity $(13)$ can be written alternatively as

$$\frac{d}{dt} \xi(t) = w + u_w = w' + u_{w'}, \tag{17}$$

and the standard rule of adding "non- relativistic" velocities takes the form

$$u_w = [w' - w] + u_{w'}. \tag{18}$$

The above definitions, given for a single word – line $(8)$, can be naturally generalized to the case of the four – velocity field defined on the whole Galilean space – time. Such field, which is a function from $G$ into $W$ (see $(4)$), shall be denoted $c(g)$. Obviously, the relations between the four – velocity field $c(g)$, the four – velocities of inertial observers $w$ and $w'$, and the "Euclidean" velocities, corresponding to the observations of the inertial observers $w$ and $w'$, are

$$c(g) = w + u_w(g) = w' + u_{w'}(g). \tag{19}$$

It is convenient to introduce inertial coordinate systems $(7)$ in a form, corresponding to the explicit choice of the „spatial" orthonormal basis $e_1, e_2, e_3$:

$$R^4 \ni (t, x_1, x_2, x_3) \to g_0 + tw + x^i e_i \in H_t \subset G. \tag{20}$$

Orthonormal basis $e_1, e_2, e_3$ in $S$ is defined by the condition

$$(e_i, e_j) = \delta_{ij}. \tag{21}$$

In $(21)$, $(...,...)$ is the alternative notation for the scalar product in $S$, which is denoted as „•"
in the definition of the structure of the Galilean space - time

$$G = (G, T_G, \gamma, \bullet). \tag{22}$$

According to the general procedure described, for example, by Wintgen and Sulanke ([3]), for
any coordinate system one can define the fields of vectors tangent to the coordinate lines and
the corresponding form fields, related by the „duality conditions". However, the important
additional property of our space is that the vector fields tangent to the coordinate lines in a
finite – dimensional affine space can be defined be means of the „affine differential quotient".
This property is related to the existence of a canonical complete derivative in the finite –
dimensional affine spaces. These properties shall be described in detail in a separate paper,
see also [6]. Now, let us compute the vector fields, tangent to the coordinate lines of the
coordinate system $(20)$. We can start with the coordinate line $x^1$; for convenience, let us
introduce the index $\alpha$, taking the values $2$ and $3$. Then

$$\frac{\partial}{\partial x^1} \square \lim_{x^{1'} \to x^1} \frac{[g_0 + tw + x^{1'}e_1 + x^\alpha e_\alpha] - [g_0 + tw + x^1 e_1 + x^\alpha e_\alpha]}{x^{1'} - x^1} =$$
$$\lim_{x^{1'} \to x^1} \frac{x^{1'} e_1 - x^1 e_1}{x^{1'} - x^1} = e_1, \tag{23}$$

and similarly

$$\frac{\partial}{\partial x^2} = e_2, \tag{24}$$

$$\frac{\partial}{\partial x^3} = e_3, \tag{25}$$

$$\frac{\partial}{\partial t} = w. \tag{26}$$

It is worth to stress that $(23), (24), (25)$ are spatial but $(26)$ is not; $(26)$ belongs to the set of
four – velocities $W$, which is a subset of $T_G$, defined by the condition $(4)$.

The vector fields, tangent to the coordinate lines of the affine coordinate systems, can be
identified with the corresponding vectors. Therefore, the fields of base forms can be also
identified with the forms from $T_G^*$ (that is, with the functionals on the translation space $T_G$ of
the Galilean space – time $G$). The corresponding "duality conditions" are

$$\left\langle Dx^i, \frac{\partial}{\partial x^j} \right\rangle = \left\langle Dx^i, e_j \right\rangle = \delta_{ij}, \tag{27}$$

$$\left\langle Dx^i, \frac{\partial}{\partial t} \right\rangle = \left\langle Dx^i, w \right\rangle = 0, \tag{28}$$

$$< Dt, \frac{\partial}{\partial x^j} > = < Dt, e_j > = 0, \tag{29}$$

$$\left\langle Dt, \frac{\partial}{\partial t} \right\rangle = \langle Dt, w \rangle = 1, \tag{30}$$

(see, for example, Wintgen and Sulanke [3]). In the above expressions, big letters denote the „four – dimensional" differentials.

For the arbitrary vector field on $G$ taking the values in $T_G$, one can introduce the corresponding complete derivative. Le us consider the vector field $V(g)$ described in a basis $(w, e_1, e_2, e_3)$ in $T_G$ (this basis corresponds to the coordinate system $(20)$ on $G$):

$$V(g) = V^0(g)w + V^i(g)e_i. \tag{31}$$

The „four – dimensional" complete derivative of $V(g)$ can be computed from the formula

$$DV(g) = w \otimes DV^0(g) + e_i \otimes DV^i(g). \tag{32}$$

The above formula results from the fact that the complete derivative in the finite – dimensional affine space, computed in the affine coordinate system, is expressed by the differentials of the coordinates. In turn, these differentials can defined by means of the "partial derivatives" (see [3]):

$$DV^0(g) = DV^0(g_0 + tw + x^i e_i) = \frac{\partial}{\partial t}V^0(g_0 + tw + x^i e_i)Dt + \frac{\partial}{\partial x^1}V^0(g_0 + tw + x^i e_i)Dx^1 +$$
$$\frac{\partial}{\partial x^2}V^0(g_0 + tw + x^i e_i)Dx^2 + \frac{\partial}{\partial x^3}V^0(g_0 + tw + x^i e_i)Dx^3 \tag{33},$$

and

$$DV^j(g) = DV^j(g_0 + tw + x^i e_i) = \frac{\partial}{\partial t}V^j(g_0 + tw + x^i e_i)Dt + \frac{\partial}{\partial x^1}V^j(g_0 + tw + x^i e_i)Dx^1 +$$
$$\frac{\partial}{\partial x^2}V^j(g_0 + tw + x^i e_i)Dx^2 + \frac{\partial}{\partial x^3}V^j(g_0 + tw + x^i e_i)Dx^3 \tag{34}$$

while these „partial derivatives" can be expressed by the „affine differential quotients", that is

$$\frac{\partial}{\partial t}V^0(g_0 + tw + x^i e_i) = \lim_{t' \to t}\frac{[V^0(g_0 + t'w + x^i e_i)] - [V^0(g_0 + tw + x^i e_i)]}{t' - t}, \tag{35}$$

etc. As it has been already mentioned in the Abstract, the limit, standing in $(35)$, does not depend on the choice of a norm in $T_G$.

Obviously, $DV(g)$ defines a field of the two – point tensor on $G$, with the first index in $T_G$ and the second index belonging to $T_G^*$. Therefore, both indices can be contracted; this contraction shall be called the „four – dimensional" divergence of a vector field on $G$:

$$DivV(g) = TrDV(g) = Tr[w \otimes DV^0(g) + e_i \otimes DV^i(g)] =$$

$$= Tr[w \otimes DV^0(g)] + Tr[e_i \otimes DV^i(g)] = \langle Dt, w \rangle \frac{\partial}{\partial t} V^0(g) + \langle Dx^j, w \rangle \frac{\partial}{\partial x^j} V^0(g) +$$

$$\langle Dt, e_i \rangle \frac{\partial}{\partial t} V^i(g) + \langle Dx^j, e_i \rangle \frac{\partial}{\partial x^j} \qquad (36)$$

In order to compute the explicit form of $(36)$ one should make use of the „duality conditions" $(27) - (30)$, and then

$$DivV(g) = \frac{\partial V(g)}{\partial t} + \frac{\partial V^i(g)}{\partial x^i}. \qquad (37)$$

A particular case of the vector field on $G$ taking the values in $T_G$ is given by the four – velocity field $(19)$. For the four – velocity field, its "four – dimensional" divergence shall be denoted as $Div\{c(g)\}$ since in our notation the four – velocity is denoted by a small letter and therefore $Divc(g)$ would be inconvenient from the "mnemonic" point of view.

In order to compute the explicit value of $Div\{c(g)\}$ one can write $(19)$ in the form that uses the orthonormal basis $e_1, e_2, e_3$ in $S$:

$$c(g) = w + u_w^i(g) e_i \qquad (38)$$

with the result that

$$Div\{c(g)\} = \frac{\partial u_w^i(g)}{\partial x^i}. \qquad (39)$$

In the standard notation, one does not write the lower index „$w$" and then $(39)$ is identical to the „Euclidean' divergence of the velocity field.

Now we shall show that the „substantial derivative" of fluid mechanics can be expressed as the „directional derivative" along the four – velocity field. The explicit calculations shall be done for the case of a real function on the space – time (denoted by $\rho(g)$) and for the four – velocity field $c(g)$ since these objects occur in the field equations for fluids.

Let us compute the complete derivative of $\rho(g)$ first:

$$D\rho(g) = D\rho(g_0 + tw + x^i e_i) = \frac{\partial \rho(g_0 + tw + x^i e_i)}{\partial t} Dt + \frac{\partial \rho(g_0 + tw + x^i e_i)}{\partial x^j} Dx^j \qquad (40)$$

where the „partial derivatives" in $(40)$ are defined by means of the corresponding „differential quotients".

In order to define the corresponding directional derivative one can either use the notation of L.Schwartz ($[6]$) or apply the notation convention taken from the tensor calculus, where the trace operation is frequently denoted by „$Tr$" while the trace operation taken with respect to the nearest indices has its own notation "$\circ$".

In turn, in mechanics of continua the substantial derivative is usually denoted „$\frac{D}{Dt}$"; as we shall see, the following identity takes place

$$\frac{D}{Dt}\rho(g) = Tr\{c(g) \otimes D\rho(g)\} = \{c(g) \circ D\rho(g)\}. \qquad (41)$$

The explicit form of the substantial derivative can be computed from the definition of contraction, after taking into account the bilinearity of the contraction and the "duality conditions":

$$\langle D\rho(g), c(g) \rangle = \left\langle \frac{\partial \rho(g)}{\partial t} Dt + \frac{\partial \rho(g)}{\partial x^i} Dx^i, w + u_w^j(g)e_j \right\rangle =$$
$$\frac{\partial}{\partial t}\rho(g) + u_w^i(g)\frac{\partial \rho(g)}{\partial x^i}. \qquad (42)$$

In the notation of Schwartz [6], the above identity takes the form

$$D_c \rho(g) = \frac{D}{Dt}\rho(g) = \frac{\partial}{\partial t}\rho(g) + u_w^i(g)\frac{\partial \rho(g)}{\partial x^i}. \qquad (43)$$

For the arbitrary point $g$ of the Galilean space – time $G$, the space $H_{[g]}$ (that is, the set of events simultaneous with the event $g$) can be parameterized by means of the following "three – dimensional" coordinates

$$(x_1, x_2, x_3) \rightarrow g + x^i e_i \in H_{[g]} \subset G \qquad (44)$$

where $e_1, e_2, e_3$ is the orthonormal basis in $S$. Also the vector fields tangent to the coordinate lines of the coordinate system $(44)$ can be computed by means of the „affine differential quotient" and it can be easily checked that they are identical to the corresponding vectors from the orthonormal basis. It is well – known that $H_{[g]}$ is an Euclidean point space and therefore the form fields, satisfying the condition

$$< dx^i, e_j > = \delta_{ij} \qquad (45)$$

can be isomorphically represented by the scalar products, that is

$$dx^i = (e_i, ...). \qquad (46)$$

For every vector field $V(g)$ on $G$, taking the values in $T_G$, it is possible to compute also the "three- dimensional" complete derivative. According to the tradition, we shall denote it as $\nabla V(g)$:

$$\nabla V(g) = \nabla V(g_0 + tw + x^i e_i) = \frac{\partial V(g_0 + tw + x^i e_i)}{\partial x^1} \otimes dx^1 +$$
$$+ \frac{\partial V(g_0 + tw + x^i e_i)}{\partial x^2} \otimes dx^2 + \frac{\partial V(g_0 + tw + x^i e_i)}{\partial x^3} \otimes dx^3. \qquad (47)$$

Till now, we have defined the „four – dimensional" complete derivative of the vector field on $G$, and, similarly, the „three – dimensional" complete derivative of the vector field on $G$. Obviously, one can define the analogous complete derivatives for the tensor fields on $G$. It can be done along the lines given by L.Schwartz [6] and the short proofs can be written with the help of the „chronological form" (see [1,2,8]). As the particular case of such tensor fields one can take the „spatial tensors" since $S$ is a subset of $T_G$. The Cauchy stress tensor of continuum mechanics is the tensor field of that kind and in a given orthonormal basis in $S$ it can be written as

$$T(g) = T^{ij}(g) e_i \otimes e_j. \tag{48}$$

The Cauchy stress tensor $(48)$ can depend, for example, on the mass density $\rho(g)$, the temperature $\Theta(g)$ and on the spatial two – point tensor $\nabla c(g)$, defined as the „three – dimensional" complete derivative of the four – velocity field $c(g)$:

$$T(g) = T(\rho(g), \Theta(g), \nabla c(g)), \tag{49}$$

(from the fact, that the difference of the two four velocities is spatial, it follows that the first index of $\nabla c(g)$ is spatial; the second index of $\nabla c(g)$ is spatial by the definition of the "three – dimensional" complete derivative).
The dependence of $(49)$ on $\nabla c(g)$ includes the viscous effects; if

$$\nabla c(g) = 0, \tag{50}$$

then

$$c(g) = Const \in W \tag{51}$$

and one can change the inertial coordinate system to the coordinate system with the vanishing velocity. Then it is possible to formulate the difference between the "fluids" and the „deformable solid bodies" (it depends on the kind of stresses present in the stationary states). Of course, we assume the "form – invariance" of our equations with respect to the change of coordinates in the inertial atlas. Such requirement automatically ensures the invariance of the theory with respect to the automorphisms of the Galilean space –time and it is independent from the "principle of material objectivity".
In order to write the explicit form of the field equations, one has to define the „gradient of the Cauchy stress tensor". Of course, it is given as the "three – dimensional complete derivative" of $(48)$

$$\nabla T(g) = e_i \otimes e_j \otimes dT^{ij}(g) = e_i \otimes e_j \otimes \frac{\partial T^{ij}(g)}{dx^\lambda} dx^\lambda, \tag{52}$$

and the „divergence of the Cauchy stress tensor" is defined as the contraction of $(52)$ with respect to the second and third indices:

$$divT(g) = \underset{2,3}{Tr} \nabla T(g). \tag{53}$$

In practice, the stress tensor $(48)$ does not depend directly on the space – time points but it is defined as the function of the „primitive fields", and the example of it is the formula $(49)$.
In order to write the field equations, we still need the substantial derivative of the four – velocity field. In order to compute it, we have to determine the complete derivative of the four – velocity first.
Eq. $(19)$ implies the identity

$$Dc(g) = D[w + u_w(g)] = D[w' + u_{w'}(g)] = Du_w(g) = Du_{w'}(g), \tag{54}$$

and $Du_w(g)$ can be written in the orthonormal reper as

$$D[u_w^i(g)e_i] = e_i \otimes Du_w^i(g) =$$
$$e_i \otimes \frac{\partial u_w^i(g_0 + tw + x^i e_i)}{\partial t} Dt + e_i \otimes \frac{\partial u_w^i(g_0 + tw + x^i e_i)}{\partial x_1} Dx^1 +$$
$$e_i \otimes \frac{\partial u_w^i(g_0 + tw + x^i e_i)}{\partial x^2} Dx^2 + e_i \otimes \frac{\partial u_w^i(g_0 + tw + x^i e_i)}{\partial x^3} Dx^3. \tag{55}$$

Our directional derivative with respect to the four – velocity field can be computed as the contraction of $(54)$ with $c(g)$, written in the form

$$c(g) = w + u_w^i(g)e_i. \tag{56}$$

According to the duality relations $(27) - (30)$:

$$D_c c(g) = <Dc(g), c(g)> =$$
$$<\frac{\partial c(g)}{\partial t} Dt + \frac{\partial c(g)}{\partial x^i} Dx^i, w + u_w^j(g)e_j> = <\frac{\partial c(g)}{\partial t} Dt, w> +$$
$$<\frac{\partial c(g)}{\partial t} Dt, u_w^j(g)e_j> + <\frac{\partial c(g)}{\partial x^i} Dx^i, w> + <\frac{\partial c(g)}{\partial x^i} Dx^i, u_w^j(g)e_j> =$$
$$\frac{\partial c(g)}{\partial t} <Dt, w> + u_w^j(g) \frac{\partial c(g)}{\partial x^i} <Dx^i, e_j> = e_i \frac{\partial u_w^i(g)}{\partial t} + e_i u_w^j(g) \frac{\partial u_w^i(g)}{\partial x^j}.$$
$$\tag{57}$$

In the standard notation, $(57)$ becomes the standard substantial derivative $\frac{D}{Dt} u^i(g)e_i$.
It can be checked that the balance of momentum can be written in the form

$$\rho D_c c(g) = divT(g) \tag{58}$$

In turn, the mass balance, the standard form of which is

$$\frac{\partial \rho}{\partial t} + \frac{\partial}{\partial x^i}(\rho u^i) = 0, \qquad (59)$$

can be equivalently written as

$$D_c \rho + \rho Div\{c\} = 0. \qquad (60)$$

We still need "the equation for the temperature". One can take it in the form of Rymarz [9]

$$\rho D_c E = T : \nabla c - \nabla q; \qquad (61)$$

the symbol „:" in (61) denotes the double contraction. For simplicity, we assume that the Cauchy stress tensor is symmetric. The other form of the "equation for the temperature" is given on p.78 of [5]. In this case, the balance of mass is still (60), in the balance of momentum the Cauchy stress tensor contains only the part determined by pressure and the "equation for the temperature" contains the entropy and is of the form:

$$D_c S = 0, \qquad (62)$$

while the pressure depends on the entropy and the mass density

$$p = p(\rho, S) \qquad (63)$$

and (63) satisfies the condition

$$v^2 = \frac{\partial p}{\partial \rho} > 0. \qquad (64)$$

One often thinks that the „entropy language" is equivalent to the „energy language". It is true for the non – relativistic ideal gases which does not exhibit quantum effects (the case of quantum gases is interesting because they can be – in a sense – "ideal", that is, their internal energy can be determined by the kinetic energy but they are a bit different from the "classical" ideal gases, and this aspect deserves a separate discussion, compare [7]).
We shall try to formulate the notion of the "adiabatic process" in the energy language. The condition (62) defines the temperature as the function of the density

$$T = T(\rho). \qquad (65)$$

It would be interesting to check, under what conditions it is possible to determine a similar function from the „energy language". After eliminating viscosity and the heat conductivity, (61) implies that

$$\rho D_c E = -p Div\{c\} \qquad (66)$$

One can insert to $(66)$ the identity

$$Div\{c\} = -\frac{1}{\rho} D_c \rho \qquad (67)$$

and obtain

$$\rho D_c E = \frac{p}{\rho} D_c \rho \qquad (68)$$

what is equivalent to

$$\frac{\rho}{p} D_c E = \frac{1}{\rho} D_c \rho \qquad (69)$$

Let us assume that there exists the constant $\alpha, \alpha > 0$, which is such that

$$\frac{\alpha p}{\rho} = E. \qquad (70)$$

For the systems with the internal energy determined by the kinetic energy, that is, for the „ideal gases", the value of this constant is explicitly known. In general, for example for dense fluids, the coefficient $\alpha$ in $(70)$ can be determined experimentally. The relation $(70)$ is motivated by the observation that it implies the relation of the form $(65)$ from the "energy language".
Now, $(70)$ implies

$$\frac{\rho}{\alpha p} = \frac{1}{E}$$

and therefore

$$\frac{\rho}{p} = \alpha \frac{1}{E}. \qquad (71)$$

After inserting $(71)$ into $(69)$ one obtains

$$\alpha \frac{1}{E} D_c E = \frac{1}{\rho} D_c \rho \qquad (72)$$

what, in turn, is equivalent to

$$\alpha D_c \ln E = D_c \ln \rho. \qquad (73)$$

From the linearity of the substantial derivative one can see that

$$D_c \alpha \ln E = D_c \ln \rho, \tag{74}$$

therefore

$$D_c \ln E^\alpha = D_c \ln \rho,$$

$$D_c \ln E^\alpha - D_c \ln \rho = 0,$$

$$D_c [\ln E^\alpha - \ln \rho] = 0,$$

and finally

$$D_c \ln \frac{E^\alpha}{\rho} = 0. \tag{75}$$

But from the Eq. $(75)$ it is possible to determine the local relation between the temperature and the density (similarly as it was in the case of the „entropy picture").
A similar reasoning can be done also in the case of mixtures, under the condition that the medium exhibits only one velocity. Then, the pressure is

$$p = p(\rho_1, \rho_2, \ldots, \rho_k, \Theta), \tag{76}$$

the density of the internal energy is

$$E = E(\rho_1, \ldots, \rho_k, \Theta), \tag{77}$$

and the Cauchy stress tensor is

$$T = T(\rho_1, \ldots, \rho_k, \Theta, \nabla c(g)). \tag{78}$$

One needs also the expression for the heat flux.
For the explicit calculations it is convenient to write $(76)-(78)$ in a different form, which takes into account that the density of the total mass is the sum of partial densities:

$$\rho = \sum_{i=1}^{i=k} \rho_i. \tag{79}$$

However, a discussion of this problem is outside the scope of the present text.

Acknowledgement. I am grateful to prof.J.Sławianowski for his help.